\documentclass[a4paper]{jpconf}
\usepackage{graphicx}
\usepackage{amsfonts}
\usepackage{amsmath}
\usepackage{amssymb}
\usepackage{amsthm}

\usepackage{float,picinpar,multicol}

\usepackage{color}
\definecolor{ks}{RGB}{0,128,0}
\definecolor{ad}{RGB}{0,255,0}
\definecolor{chi}{RGB}{255,200,0}
\definecolor{likelihood}{RGB}{255,0,0}

\usepackage{type1cm}
\usepackage{eso-pic}
\usepackage{epstopdf}

\begin{document}
\title{New Statistical Techniques in the Measurement of the inclusive Top Pair Production Cross Section}
\author{ Ji\v{r}\'{i} Franc, Petr Bou\v{r}, Michal \v{S}t\v{e}p\'{a}nek, and V\'{a}clav K\r{u}s}
\address{Department of Mathematics, Faculty of Nuclear Sciences and Physical Engineering,
Czech Technical University, Trojanova 13, 120 00 Prague 2, The Czech Republic}
\ead{jiri.franc@fjfi.cvut.cz, bourpetr@fjfi.cvut.cz, michal.stepanek@fjfi.cvut.cz, vaclav.kus@fjfi.cvut.cz}

\begin{abstract}
We present several different types of multivariate statistical techniques used in the measurement of the inclusive top pair production cross section in $p \bar{p}$-collisions at $\sqrt{s} = 1.96 \text{TeV}$ employing the full RunII data  ($9.7\textrm{ fb}^{-1}$) collected with the D0 detector at the Fermilab Tevatron Collider. We consider the final state of the top quark pair decays containing one electron or muon and at least two jets. We proceed various statistical homogeneity tests such as Anderson - Darling, Kolmogorov - Smirnov, and $\varphi$-divergences tests to determine, which variables have good data-MC agreement, as well as a good separation power. We adjusted all tests for using weighted empirical distribution functions. Further we separate $t\bar{t}$ signal from the background by the application of Generalized Linear Models, Gaussian Mixture Models, Neural Networks with Switching Units and confront them with familiar methods from ROOT TMVA package such as Boosted Decision Trees, and Multi-layer Perceptron. We compare results by area under receiver operating characteristic curve and verify the quality of the discrimination from all methods. 
\end{abstract}


\section{Introduction}
 
The main goal of this analysis is to apply statistical techniques, which are new to HEP, to the measurement of the inclusive top pair production cross section in $p \bar{p}$-collisions at $\sqrt{s} = 1.96 \text{TeV}$ employing the full D0 RunII data. Top quarks are identified by their characteristic decay into the lepton+jets final state. Background events with a similar final state originate from other physics processes or from misidentification. The multivariate methods discussed aim to separate signal and background.\\
Since we do not know, if recorded data events from the detector belong to signal or background, we have to train our classification methods on Monte Carlo samples that simulate the situation in the detector in terms of the standard model of particle physics. The generated MC sample is splitted into three groups. One half is used for training and testing (50\% respectively) and the other half (called yield) is used for the application and final analysis with data. The final MC selection with the description of all applied kinematic range and additional quality cuts is described in \cite{Selection}. From approximately 50 different available kinematical and topological variables we selected 30 variables showing a good data/MC agreement as well as a good separation power.\\
Classical approach of the selection of variables is based on visual checks so-called controlplots, where the histograms of both samples, data and MC, are figured in one subplot. Additional verification is done trough Kolmogorov - Smirnov test applied on mentioned histograms with certain number of bins, without any information on original empirical distribution functions. We applied more rigorous statistical approach and we modified common homogeneity tests, in terms of adding weights and utilization of quantile binning.

\section{Selection of Variables and Homogeneity Tests}  

In order to perform the efficient training of separation methods on MC simulation we need to guarantee the homogeneity of both MC and data populations, i.e., we test the following hypothesis:
\begin{equation}\label{eq:hypo}
	H_0: F=G \qquad \textrm{versus} \qquad H_1: F \neq G \qquad \textrm{at significance level $\alpha$},
\end{equation}
\noindent where $F$ is unknown cumulative distribution function (CDF) of data distribution and $G$ is unknown CDF of MC distribution. Let $\boldsymbol{X}_1=\{X_1,\ldots,X_{n_1}\}$ denote a random sample taken from the distribution $F$ and $\boldsymbol{X}_2=\{Y_1,\ldots,Y_{n_2}\}$ be a random sample taken from the distribution $G$. We further denote $N=n_1+n_2$. Let $F_{n_1},G_{n_2}$ denote empirical distribution functions (EDFs) of samples $\boldsymbol{X}_1,\boldsymbol{X}_2$ respectively. In our hypothesis testing we seek for \emph{p}-value, i.e., the lowest significance level $\alpha$ at which we reject $H_0$. Thus, we automatically reject $H_0$ for every higher significance level $\alpha >p\textrm{-value}$.

However, MC simulation is weighted by weights $(w_1,\ldots,w_{n_2})$, so we process the MC sample $\boldsymbol{X}_2^w=\{(Y_1,w_1),\ldots,(Y_{n_2},w_{n_2})\}$. Therefore, we are forced to replace EDF with weighted empirical cumulative distribution function (WEDF) defined by $G_{n_2}^w(x)=\frac{1}{W_2}\sum_{i=1}^{n_2}{w_i\boldsymbol{1}_{\left(-\infty,x\right]}(x_i)}$, where $W_2=\sum_{i=1}^{n_2}{w_i}$ and $\boldsymbol{1}_{\left(-\infty,x\right] }$ is an indicator function of the set $\left(-\infty,x\right]$. Since all the weights are equal to 1 in data sample, the WEDF $F_{n_1}^w$ coincides with the EDF $F_{n_1}$. Let us denote $W=W_1+W_2$, where $W_1 = \sum_{j=1}^{n_1}{w_j}$ for the case of $F_{n_1}^w$.

\subsection{Divergence Tests of Homogeneity}
This test reduces the problem \eqref{eq:hypo} to testing homogeneity in multinomial populations. Let $\{t_1,\ldots,t_{m+1}\}$ be a partition of the real line such that for all $x\in\{\boldsymbol{X}_1,\boldsymbol{X}_2\}$ it holds that $x\in\left[ t_1,t_{m+1}\right]$. Hereby, we make this binning over populations $\boldsymbol{X}_1,\boldsymbol{X}_2$ consisting of $m$ bins. For $i \in \{ 1,2 \}$ and $j \in \{ 1,\ldots,m \}$ we denote by $p_{ij}$  the probability that a randomly chosen observation from $\boldsymbol{X}_i$ belongs to the $j$-th bin. Instead of \eqref{eq:hypo} we now test hypotheses
\begin{equation}
	H_0: p_{1j}=p_{2j} \quad \forall \quad j\in \{1,\ldots,m\} \qquad \textrm{versus} \qquad H_1: H_0 \textrm{ is not true}.
\end{equation}

\noindent As given in \cite{Pardo2006} using $\phi$-divergence measure and maximum likelihood estimators $n_{ij}/n_i$ and $N_j/N$, we consider the test statistic
\begin{equation}\label{eq:stat}
	H_N^{\phi} = \frac{2N}{\phi''(1)}\sum_{i=1}^2{\sum_{j=1}^m{\frac{n_i}{N}\frac{N_j}{N}\phi\left(\frac{n_{ij}N}{n_iN_j}\right)}},
\end{equation}

\noindent where $n_{ij}$ is the number of observations from $\boldsymbol{X}_i$ in $j$-th bin, $N_j$ is the number of all observations in $j$-th bin and $\phi$ is a given function from convex family. For our purposes, the maximum likelihood ratio estimators $n_{ij}/n_i,N_j/N$ need to be replaced with the corresponding versions of weighted estimators $w_{ij}^{\textrm{(in bin)}}/W_i,w_j^{\textrm{(in bin)}}/W$ made of the respective sums of weights. The asymptotic distribution of the test statistic \eqref{eq:stat} is $\chi^2$ with $(m-1)$ degrees of freedom. Thus, the approximate \emph{p}-value can be obtained as $1-\chi_{(m-1)}^2\left(H_N^{\phi}\right)$. There are two well-known specific cases: for $\phi(x)=\frac{1}{2}(x-1)^2$ the test coincides with the $\chi^2$ Homogeneity Test and for $\phi(x)=x\log x-x+1$ the test is identically the Likelihood Ratio Test.

\subsection{Kolmogorov-Smirnov Test for Two Samples}
Unlike $\chi^2$ test, Kolmogorov-Smirnov test is based on differences between EDFs of two samples. We consider the statistic $D_{n_1,n_2}=\sup_{x\in\mathbb{R}}|F_{n_1}(x)-G_{n_2}(x)|$. It follows from the Glivenko-Cantelli lemma that under the true $H_0$ it holds that $D_{n_1,n_2} \stackrel{a.s.}{\to} 0$ for $n_1,n_2 \to \infty$. Furthermore, due to \cite{Smirnov1944} it holds for the true $H_0$ and $\lambda>0$
\begin{equation}\label{eq:kolm}
	\lim_{n_1,n_2\to\infty}{P\left(\sqrt{\frac{n_1n_2}{n_1+n_2}}D_{n_1,n_2}\leq\lambda\right)}=1-2\sum_{k=1}^{\infty}{(-1)^{k-1}e^{-2k^2\lambda^2}}.
\end{equation}

\noindent Therefore, we can obtain approximate \emph{p}-value as $2\sum_{k=1}^{\infty}{(-1)^{k-1}e^{-2k^2\lambda_0^2}}$, where $\lambda_0=\sqrt{\frac{n_1n_2}{n_1+n_2}}D_{n_1,n_2}$. In \eqref{eq:kolm}, the EDFs and $n_1,n_2$ need to be replaced with the corresponding WEDFs and sums of weights $W_1,W_2$, respectively. Kolmogorov-Smirnov test is more powerful than $\chi^2$ test generally , see \cite{Stephens1992}.

\subsection{Anderson-Darling Test for Two Samples}
Another test based on EDF is Anderson-Darling test. Here $H_N$ stands for EDF of pooled sample $\{\boldsymbol{X}_1,\boldsymbol{X}_2\}$, i.e., $H_N(x)=\left[n_1F_{n_1}(x)+n_2G_{n_2}(x)\right]/N$. We take into account the statistic from \cite{Pettitt1976}
\begin{equation}\label{eq:stdstat}
	A_{n_1n_2}^2=\frac{n_1n_2}{N}\int_{-\infty}^{+\infty}{\frac{\left[F_{n_1}(x)-G_{n_2}(x)\right]^2}{H_N(x)\left[1-H_N(x)\right]}\textrm{d}H_N(x)},\qquad T_{n_1n_2}=\frac{A_{n_1n_2}^2-1}{\sigma_N},
\end{equation}

\noindent where $\sigma_N=\textrm{var}\left(A_{n_1n_2}^2\right)$. According to \cite{Scholz1987} we can determine an approximate \emph{p}-value by means of the standardized statistic $T_{n_1n_2}$. Once again,  in \eqref{eq:stdstat} we need to replace the EDFs and the numbers of entries $n_1,n_2,N$ with their respective WEDFs and sums of weights $W_1,W_2,W$. Anderson-Darling test is generally more powerful than Kolmogorov-Smirnov test, see \cite{Engmann2011} in detail.

\subsection{Results}
We tested all 50 potential input variables and we made a selection with 36 of them. Here we present a preview of results for some of them in electron channel.
\begin{figure}[H]
\centering
%
%
%
%
%
%
\includegraphics[width=15cm]{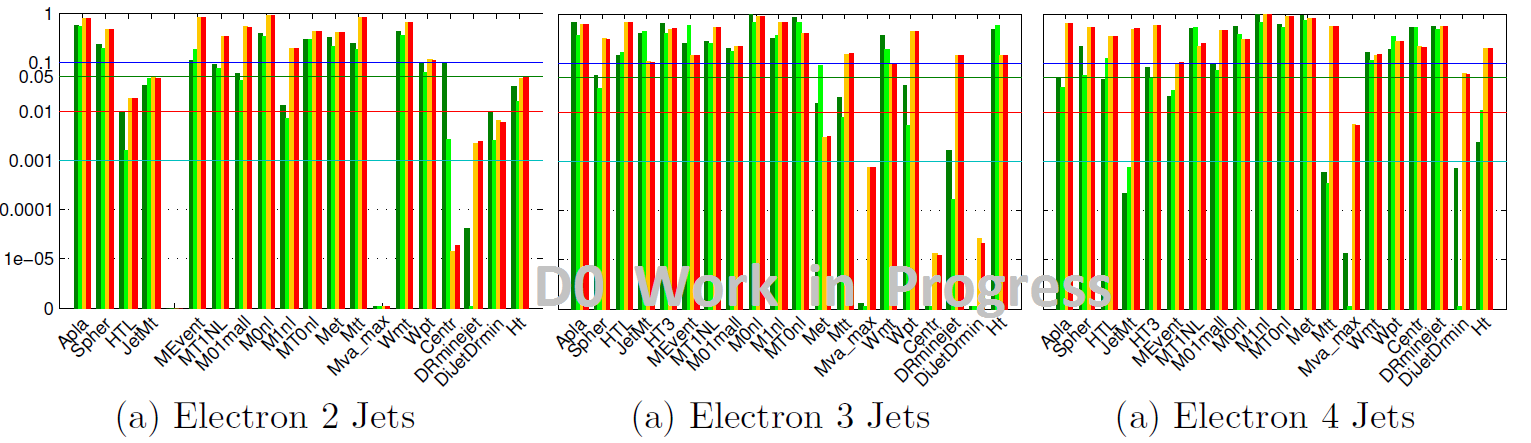}
\caption{Approximate \emph{p}-values for selected variables from different tests MC vs data:\\
\textcolor{ks}{$\blacksquare$} Kolmogorov-Smirnov, \textcolor{ad}{$\blacksquare$} Anderson-Darling, \textcolor{chi}{$\blacksquare$} $\chi^2$, \textcolor{likelihood}{$\blacksquare$} Likelihood Ratio.\\
(variable \textsf{HT3} is not available for channel with only 2 Jets)}
\end{figure}

\section{Discrimination}
\begin{figure}[h!tbp]

\centering
             \includegraphics[width=15cm]{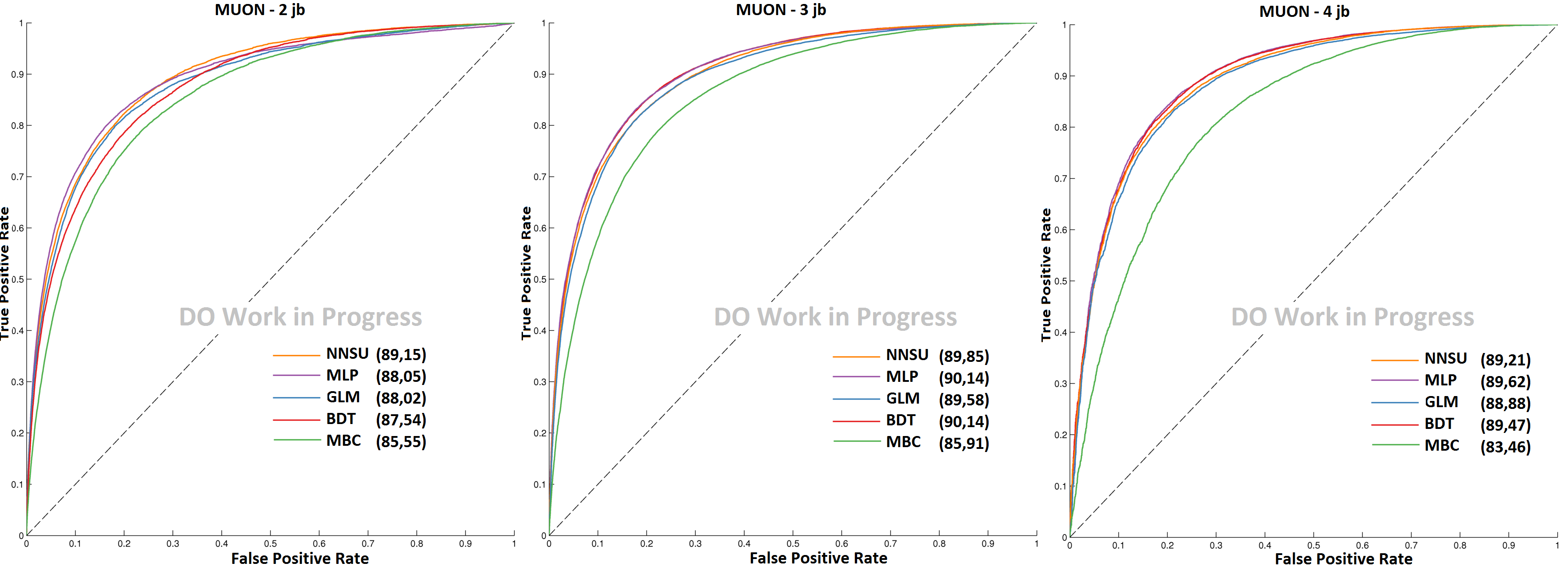}
        	\caption{ROC curves for different methods and analysis chanels}
        	\label{results2}
  \end{figure}

For a long time till the end of the last century, the High Energy Physiscs (HEP) community used for the discrimination linear decision boundary methods (Fisher-discriminants) and later Naive Bayesian methods. Not until the first decade of the 21st century the supervised machine learning methods such as Neural Networks, Boosted Decision Trees, and Support Vector Machines began to play important role in the HEP analysis. Nowadays, the multivariate analysis techniques are one of the fundamental tools in the discrimination phase. Nevertheless, there are still some well known statistical methods that are worth trying out. Let us mention three methods, whose quality of separation was tested in the measurement of the inclusive top pair production cross section on D0 Tevatron full RunII data. The first one is  Model Based Clustering method (MBC) based on EM algorithm and Gaussian Mixture Models presented in \cite{TOP2013}, the second one is well-known Generalized Linear Models (GLM), where we tested different link functions and overdispersions, and the third one is Neural Nets with Switching Units (NNSU, quite new method developed by F. Hakl from Institute of Computer Science of the ASCR).\\
We compared mentioned approaches with another MVA methods such as Multilayer Perceptron (MLP) and Boosted Decisions Trees (BDT) from the ROOT TMVA package. ROC curves as an example for all three muon + jets bin are shown in Figure \ref{results2}. It's clear that new methods are comparable and provide similar results as established methods from the TMVA package.

%

\section{Discussion}
We presented the generalization of statistical homogeneity tests and their utilization in HEP analysis. Since in the ROOT framework the  most used classical K-S test is designed for histograms instead of weighted empirical function,  our approach is more proper. In a similar way we modified Anderson-Darling test that is more powerful than the two sample K-S test. However, both mentioned tests are very sensitive and that's why we recommend using $\varphi$-divergence tests of homogeneity with quantile binning, where the utilization of weights is more straightforward. Depending on the used convex function $\varphi$ we can convert the test to the $\chi^2$-test, Likelihood Ratio test, or any other general $\varphi$-divergence tests.
Further, we shortly mentioned not so common by used discrimination methods, whose quality of the signal from background separation is comparable with methods from ROOT TMVA package.

\vspace{-1mm}
\ack{This work has been supported by the MSMT (CZ) grant INGO II INFRA LG12020 and CTU (CZ) grant SGS12/197/OHK4/3T/14.}

\end{document}